\documentstyle{article}
\renewcommand{\thefootnote}{\fnsymbol{footnote}}

\begin{document}

\begin{center}
{\Large \bf Algebraic Aspects of Interactions \\
of Massive Spinning Particles In Three Dimensions } \\

\vspace{4mm}

K.B. Alkalaev \footnote{E-mail: alkalaev@phys.tsu.ru} and
S.L.  Lyakhovich \footnote{E-mail: sll@phys.tsu.ru} \\ Department of
Physics, Tomsk State University, 634050 Tomsk, Russia \\

\end{center}
\vspace{0.4cm}
\begin{center}
Talk given at XIV  International Workshop on High Energy Physics and
Quantum Field Theory, QFTHEP'99 \\
Moscow, Russia, May 27- June 2, 1999
\end{center}
\vspace{0.4cm}

\begin{abstract}
The most general 2+1 dimensional spinning particle model is considered with
the configuration space ${\cal M}^5=R^{1,2}\times{\cal L}$ where
${\cal L}$ is a Lobachevsky plane being a space of the spinning modes. The
action functional may involve all the possible first order Poincare
invariants of ${\cal M}^5$ world lines, and the particular class of actions
is specified thus the corresponding gauge algebra to be unbroken by
inhomogeneous external fields.  Nevertheless, the consistency problem reveals
itself as a requirement of the global compatibility between first and second
class constraints. These compatibility conditions, being unnoticed before in
realistic second class theories, can be satisfied for a particle iff the
gyromagnetic ratio takes the critical value $g=2$. The quantization procedure
is suggested for a particle in the generic background field by making use of
a Darboux co-ordinates, being found by a perturbative expansion in the field
multipoles and the general procedure is described for constructing of the
respective transformation in any order.

\end{abstract}

%-----------------------------------------------------------------------------
\section{Introduction}

\setcounter{equation}{0}
%-----------------------------------------------------------------------------

It is always a problem to construct consistent interactions for systems which
involve higher spins. The problem appears in any dimension both at level of
particles, fields and strings. For the massless case in low dimensions, the
interacting spin fields allow for a uniform description at the level of
equations of motion in the AdS space \cite{FradkinVasiliev,Vasiliev}.
Free massive field actions have long been known, it is not the
case for interactions. When an action possesses invariance under higher-spin
transformations, the reason of inconsistency is well known - the gauge
algebra may be broken down when the interaction deforms theory.
However, the peculiar feature of massive higher spin
fields is that they may be formulated as pure second class constrained
theories having no gauge invariance from the very beginning. The example of
gauge noninvariant action was given in Ref. \cite{SH} for the massive higher
spin fields.  In this case, the consistency criteria become less obvious,
however one should require at least Poincare invariance, renormalizability,
unitarity and possibly somewhat yet unknown.  Using the
action of Ref.  \cite{SH} the prescription was given in Refs. \cite{FPT,P} for
imposing tree-level unitarity condition on a whole energy spectrum of the
Compton scattering amplitudes of a single massive higher spin field. Some of
coefficients in the action were fixed unambiguously, in particular, in a
homogeneous e/m field gyromagnetic ratio takes the value $g=2$ \cite{FPT}.

Another way to get consistent interacting field equations is to quantize
an appropriate classical mechanical model of massive spinning
particle which is constructed as a constrained Hamiltonian system with a
finite number degrees of freedom.  The advantage of the method is that it is
simpler to achieve consistent couplings to backgrounds at the classical level
than to examine it in the quantum field theory. The reason is that there is
only a finite number of gauge symmetries ( first class constraints ) in the
classical particle description and continuous infinity of field theoretical
symmetries are generated from them through a quantization procedure. In
Ref.  \cite{universal model} the systematic procedure was developed to
construct the most general action functional ( involving some external
parameters ) for a particle of an arbitrary spin. One may restrict
parameter's values by appropriate conditions to the special case which admits
consistent interactions with a general background. However, as further will
be clarified, the consistency problem is not exhausted by noncontradictory
deformation of the gauge algebra.  It appears again as a requirement of
global smoothness of equations of motion on the phase space.  More precisely,
when the interaction is switching on, the special surface appears where the
Dirac brackets become singular, and to make equations of motion regular upon
this surface we should to impose some smoothness conditions restricting the
choice of the free parameters in the action.
These conditions, being applied for the case of the homogeneous e/m field,
fix the gyromagnetic ratio as $g=2$ which is the same value that
has been deduced from the field theoretical approach \cite{FPT}.

The note is organized as follows. We begin in section 2 with a brief view of
the universal model of 3D particle, being the 3D analog of the model,
allowing for the consistent interaction in D=4 \cite{universal model}.
In section 3 the problem
of interaction is considered, the singular structure of the phase space is
clarified, and the consistency conditions are obtained. In section 4 we turn
to quantization of the model and in section 5 we conclude with some general
remarks on the perturbative quantization procedure and the global
consistency problem. The Appendix is attached with the some details of the
geometric constructions.

%-----------------------------------------------------------------------------
\section{Universal 3D spinning particle model}

\setcounter{footnote}{0}
\renewcommand{\thefootnote}{\alph{footnote}}
\setcounter{equation}{0}
%-----------------------------------------------------------------------------

Configuration space of 3D spinning particle \cite{anyon,superanyon} can be
chosen as ${\cal M}^5=R^{1,2}\times{\cal L}$, where the Lobachevsky plane
${\cal L}$ carries spinning degree of freedom \footnote{For the details of
inner space geometry and notations to be used in the text see the Appendix,
more details of this geometry could be found in \cite{anyon}. }.  Poincare
generators
\begin{equation}
P_{a}=p_{a}
\qquad
J_{a}=\varepsilon_{abc}x^{b}p^{c}+j_{a}
\end{equation}
being functions on the
extended phase space $T^{*}{\cal M}^{5}$ form the algebra with respect to the
Poisson brackets
\begin{equation}
\begin{array}{rcl}
\{ P_{a},P_{b} \}& = & 0 \\
\{ P_{a},J_{b} \} & = & \varepsilon_{abc}\eta^{cd}P_{d} \\
\{ J_{a},J_{b} \} & = & \varepsilon_{abc}\eta^{cd}J_{d}
\end{array}
\end{equation}
Level surface of Casimir functions $P^{2}=-m^{2}$ and
$PJ=ms$ fixes  particle's spin and mass
parameters in the space of values $(P_{a},J_{a})$. Due to commutations
relations (2.2) Hamiltonian counterparts for the Casimirs on the extended
phase space should appear as first class constraints.  Existence of two
independent gauge symmetries reveals itself in the form of classical
Zitterbewegung. When the effect is presented the particle's world-surface is
no more straight line but takes topology of cylinder  \cite{anyon}.
The same phenomenon may happened in not only in d=3, but in d=4
\cite{universal model} and higher dimensional case \cite{hdim1}.  To reduce
extra dimensions of the world-surface one should require the
gauge symmetries to be dependent in space-time part of extended phase space
\begin{equation}
\left\{ \begin{array}{rcl}
\delta_{1\varepsilon}x^{a}&=&2\varepsilon p^{a} \\
\delta_{1\varepsilon} p^{a}&=&0
\end{array}
\right\}
\sim
\left\{
\begin{array}{rcl}
\delta_{2\mu} x^{a}&=&\mu j^{a} \\
\delta_{2\mu} p^{a}&=&0
\end{array}
\right\}
\end{equation}
Taking account of Casimirs surfaces,
the requirement results in $p_{a}=-m/s j_{a}$.

To construct the most general action functional \cite{universal model}
we need to subject it to the following natural conditions:

 1) the Lagrangian does not contain higher derivatives

 2) the action is invariant under reparametrization of particle's world-line

 3) mass-shell $P^{2}+m^{2}=0$ and spin-shell $PJ-ms=0$ conditions should
arise in the theory as constraints

There are five true Poincare invariants (for definition of $n_a$ and other
notation details, see Appendix)
\begin{equation}
\begin{array}{rcl}
\Gamma_{1}& = & \dot{x}^{2} \\
\Gamma_{2}& = & (\dot{x}n)^{2} \\
\Gamma_{3}& = & (\dot{x}\dot{n}) \\
\Gamma_{4}& = & \varepsilon_{abc}\dot{x}^{a}n^{b}\dot{n}^{c} \\
\Gamma_{5}& = & \dot{n}^{2}
\end{array}
\end{equation}
and one more value which transforms by a total derivative
\begin{equation}
\Gamma =  i\frac{\dot{z}\bar{z}-\dot{\bar{z}}z}{1-z\bar{z}}
\end{equation}
The most general Poincar\'e- and reparametrization-invariant action has
reads
\begin{equation}
S = \int d\tau(L(\Gamma_{i})+\beta\Gamma )\equiv\int
d\tau{\cal L}\
\end{equation}
Satisfying all mentioned conditions for the action to be consistent with
interactions one gets the one-parametric family of the Lagrangians
\begin{equation}
{\cal L}_{\gamma}=m\gamma
(\dot{x}n)+\frac{m}{s}\sqrt{(\gamma^{2}-1)(\dot{x}^{2}+(\dot{x}n)^{2}
-\frac{2s}{m}\varepsilon_{abc}\dot{x}^{a}n^{b}\dot{n}^{c}
+\frac{s^{2}}{m^{2}}\dot{n}^{2})}
+i\gamma s\frac{\dot{z}\bar{z}-\dot{\bar{z}}z}{1-z\bar{z}}
\end{equation}
where $\gamma =\beta /s$. When the parameter $\gamma\neq 1$, the Lagrangian
possesses two gauge symmetries being dependent in the space-time part of
${\cal M}^{5}$. If $\gamma =1$, these gauge symmetries become dependent
in the whole configuration space, and they both reduce to reparametrization
invariance only. In what follows we consider only the case $\gamma =1$.

%-----------------------------------------------------------------------------
\section{Interactions}

\setcounter{equation}{0}
%-----------------------------------------------------------------------------

The full description of the most general interacting Lagrangian and its
constrained Hamiltonian analysis is described in Ref. \cite{AL}. In this
section we examine the case of homogeneous e/m field:
\begin{equation} L=(\dot{x}n)(m+e\gamma(Fn))
+is\frac{\dot{z}\bar{z}-\dot{\bar{z}}z}{1-z\bar{z}}
-e\dot{x}^{a}(A_{a}+\mu F_{a}+ \delta F_{ab}n^{b})+{\cal O}(e^{2})
\end{equation}
here $\gamma , \mu$ and $\delta$ are arbitrary constants. When the field is
homogeneous, the term $-e\dot{x}^{a}\mu F_{a}$ does not contribute to
dynamics as it vanishes modulo a total time derivative. The complete set of
the constraints on ${\cal M}^{8}=T^{*}(R^{1,2})\times {\cal L}$ reads
\begin{equation}
T_{a}=p_{a}-mn_{a}-e\gamma(Fn)n_{a}+e\delta F_{ab}n^{b}
\label{ta}
\end{equation}
Their brackets on ${\cal M}^{8}$ may be presented in the form
\begin{equation}
\{T_{a},T_{b}\}=
\varepsilon_{abc}((1-\frac{me\gamma}{s})F^{c}
-(\frac{m^{2}}{s}+\frac{2me\gamma}{s}(Fn))n^{c}
+\frac{me\delta}{s}F^{ck}n_{k})
\equiv\varepsilon_{abc}N^{c}
\label{tatb}
\end{equation}
The maximal rank of this matrix equals 2 and corresponding null-vectors are
$N_{c}$. Thus, in a general position, there are two second class constraints
and one first class among of $T_{a}$.  As is obvious,
the rank decreases to zero if the r.h.s vanished in  (\ref{tatb})
$N_{c}=0$. Locally, i.e. {\it on the surface} $N_{c}=0$,
all the constraints become the "first class". The Dirac
brackets, being constructed of the second class constraints in the general
position, become singular on this surface that gives rise to the
discontinuity in the equations of motion.
To study the dynamics near the singular surface, it is convenient to chose
another constraint basis which is equivalent to the original constraints
$T_{a}=0$ (\ref{ta}), the equivalence immediately follows from the identity
(6.5),
\begin{equation}
H=p^{2}+m^{2}+2em\gamma (Fn)+e\delta F_{ab}p^{a}n^{b}
\end{equation}
\begin{equation}
\theta =(p\xi)+ie\delta(F\xi) \qquad
\bar{\theta} =(p\bar{\xi})-ie\delta(F\bar{\xi})
\end{equation}
The algebra of these constraints reads
\begin{equation}
\{H,\theta\}=-2ie(pn)(F\xi)((1-\frac{m\gamma}{s})+\frac{i\delta}{s}(pn))+
(\sim\theta ,\bar{\theta})
\end{equation}
\begin{equation}
\{H,\bar{\theta}\}=2ie(pn)(F\bar{\xi})((1-\frac{m\gamma}{s})-\frac{i\delta}{s}(pn))+
(\sim\theta ,\bar{\theta})
\end{equation}
\begin{equation}
\{\theta,\bar{\theta}\}=-\frac{i\zeta^{2}}{2s}((pn)^{2}+s(Fn))+(\sim
\theta ,\bar{\theta})\equiv\Phi+(\sim\theta ,\bar{\theta})
\end{equation}
\begin{equation}
\Phi\equiv-\frac{i\zeta^{2}}{2s}((pn)^{2}+s(Fn))
\end{equation}
Denote constraints $(\theta ,\bar{\theta})=\theta_{i} \, , \, i=1,2$
and co-ordinates on ${\cal M}^{6}=T^{*}(R^{1,2})$ as
$(x_{a},p^{b})=\Gamma^{A}$. Then Dirac brackets take the form
\begin{equation}
\omega^{AB}=\{\Gamma^{A},\Gamma^{B}\}_{DB}=
\{\Gamma^{A},\Gamma^{B}\}+\frac{1}{\Phi}
\{\Gamma^{A},\theta_{i}\}\varepsilon_{ij}\{\theta_{j},\Gamma^{B}\}
\end{equation}
To reduce inner space, from $T_{a}=0$ we exclude vector $n_{a}$ \cite{AL} as
\begin{equation}
n_{a}=\frac{p_{a}}{(-p^{2})^{1/2}}-e\delta\frac{F_{ab}p^{b}}{p^{2}}+{\cal
O}(e^{2})
\end{equation}
One can see that tensor $\omega^{AB}$ becomes singular on the surface
\begin{equation}
\Phi\equiv p^{2}-s\frac{Fp}{(-p^{2})^{1/2}}=0
\end{equation}
Thus the phase space ${\cal M}^{6}$, being equipped with the bracket
tensor $\omega^{AB}=\{\Gamma^{A},\Gamma^{B}\}$,  is not a smooth manifold
anymore, it is decomposed into 3 parts:
${\cal M}^{6}={\cal M}^{-}\cup\Phi\cup{\cal M}^{+}$, where
${\cal M}^{-}$ and ${\cal M}^{+}$ are
domains of analiticity of the bracket tensor. These domains are
defined by the conditions $\Phi <0$ and $\Phi >0$ respectively.
The equations of motion
\begin{equation}
\dot{\Gamma}^{A}=\{\Gamma^{A},H\}+\frac{1}{\Phi}
\{\Gamma^{A},\theta_{i}\}\varepsilon_{ij}\{\theta_{j},H\}
\end{equation}
may suffer from a discontinuity at $\Phi =0$. To make the equations of motion
smooth one should restrict dynamics on physical space with the following
condition to be satisfied
\begin{equation}
\{H,\theta_{i}\}=\Pi_{ij}\theta_{j}+K_{i}H
\end{equation}
where $\Pi_{ij}$ and $K_{i}$ are arbitrary functions. When the condition
(3.14) is satisfied, the singular second term in (3.13) disappears, at
least on shell.
For the algebra (3.7-8) this means that parameters take
their critical values $\delta =0$ and $\gamma =s/m$. The equations of motion
become smooth in the whole phase space ${\cal M}^{6}$ and take the well-known
"Lorentz" form
\begin{equation}
\dot{x}^{a}=2p^{a}+{\cal O}(F^{2},\partial F,...)
\end{equation}
\begin{equation}
\dot{p}^{a}=-2eF^{ab}p_{b}+{\cal O}(F^{2},\partial F,...)
\end{equation}
\begin{equation}
H=p^{2}+m^{2}+2se\frac{(Fp)}{(-p^{2})^{1/2}}+{\cal O}(F^{2},\partial
F,...) =0
\end{equation}
So far we have considered only the first order in the charge $e$ but, in the
homogeneous field, the procedure may be closed in coupling constant. The
equations (3.15-17) remain the same and Lagrangian takes the form
\begin{equation}
L=(\dot{x}n)\sqrt{m^{2}+2se(Fn)}
+is\frac{\dot{z}\bar{z}-\dot{\bar{z}}z}{1-z\bar{z}}
-e\dot{x}^{a}A_{a}
\end{equation}
If one calls numerical coefficient at the vertex $(Fj)$ in Lagrangian
(3.18) and Hamiltonian (3.17) as a gyromagnetic ratio \cite{BMT},
then one may conclude that the smoothing condition (3.14) fixes it as
$\gamma=2$.

Consider the consistency problem (which has been treated above in the
framework of the constrained dynamics) from the viewpoint of the symplectic
geometry. Symplectic view allows to formulate the consistency condition
directly in the reduced phase space ${\cal M}^{6}$. The symplectic 2-form
associated to Dirac brackets
\begin{equation}
\Omega
=\frac{1}{2}(\omega^{-1})_{AB}d\Gamma^{A}\wedge d\Gamma^{B}
\label{omega}
\end{equation}
degenerates on the singular surface $\Phi =0$:  $\det\Omega =\Phi^{2}$.
Equations of motion have the form \begin{equation}
\Omega_{AB}\dot{\Gamma}^{B}|_{H=0}=\partial_{A}H|_{H=0}
\end{equation}
and symplectic counterpart for the condition (3.13) reads as
\begin{equation}
Z^{i}_{A}\partial^{A}H|_{\Phi=H=0} =0
\label{zh}
\end{equation}
where $Z^{i}_{A}$ are $[i]$ linear
independent {\em local} null-vectors of the two-form (\ref{omega}):
\begin{equation}
\Omega_{AB}Z^{B}_{i}|_{\Phi=0} =0
\end{equation}
The vectors $Z^{i}_{A}$ are defined
by the two-form $\Omega$ regardless to the Hamiltonian $H$. Since
$Z^{i}_{A}$ are fixed, relations (\ref{zh}) should be treated as restrictions
to the admissible form of the Hamiltonian. It is the restriction which fixes
the gyromagnetic ratio in the case of the particle in external e/m field.

%-----------------------------------------------------------------------------
\section{Quantization}

\setcounter{equation}{0}
%------------------------------------------------------------------------------

There exist a good many of different methods of quantization of a given
constrained system. Nevertheless, it is always a problematic task to quantize
when phase space possesses nonlinear Poisson brackets. Usually, the
quantization procedure is understood in the sense of deformation in Plank
constant of the classical brackets to quantum commutators whose first
order term is proportional to the classical ones. This approach is
commonly regarded as practically ineffective for the case of general
manifold because there is no representation theory for quantum brackets.
The quantization problem is strongly simplified in a moment if one finds
the canonical realization ( Darboux co-ordinates ) for the initial classical
Poisson brackets.  This formal idea takes sensible form when phase space of
interacting spinning particle is considered. The interacting, theory
contains coupling constant $e$ and knowing Darboux realization for free
particle sector one may look for Darboux transformation of initial
"interacting" co-ordinates in the form of perturbative expansion in coupling
constant $e$.  The attempt to construct the transformation in first order in
$e$ was originally taken in Ref. \cite{CNP} in the framework of a  minimal
anyon model. Below, we develop the general procedure for constructing of the
respective transformation to the Darboux coordinates in any order in charge.

The structure of the proposed perturbation approach is most manifest in terms
of symplectic geometry. In Darboux co-ordinates $(x,p)$, 2-form of free case
$\Omega_{0}=\Omega(e=0)$ takes the canonical form
\begin{equation}
\Omega_{0}=\Omega(e=0) \rightarrow \Omega=dp_{a}\wedge dx^{a}
\end{equation}
To find canonical representation for the interacting 2-form
$\Omega=\Omega(e)$ we proceed follows: realize  the
initial variables $\Gamma$ in the interacting theory
 as perturbation expansions in
$e$ of the initial variables of the free theory $\Gamma_{0}$
and then, making change of variables $\Gamma_{0}=\Gamma_{0}(x,p)$,
bring $\Omega=\Omega(e)$ to the canonical form:
\begin{equation}
\Omega_{0}=\Omega(e) \stackrel{\Gamma=\Gamma(\Gamma_{0})}{\longrightarrow}
\Omega_{0}=\Omega(e=0)
\stackrel{\Gamma_{0}=\Gamma_{0}(x,p)}{\longrightarrow} dp_{a}\wedge dx^{a}
\end{equation}
In more practical terms the mentioned procedure means the following:
starting with a pair of 2-forms, $\Omega_{0}$ for the free case and
$\Omega$ for the interacting one
\begin{equation}
\Omega
=\Omega_{AB}(\Gamma ; e)d\Gamma^{A}\wedge d\Gamma^{B}=
\sum_{n=0}^{\infty}e^{n}\Omega^{n}_{AB}(\Gamma)d\Gamma^{A}\wedge d\Gamma^{B}
\end{equation}
\begin{equation}
\Omega_{0}=\Omega^{0}_{AB}(\Gamma_{0})d\Gamma^{A}_{0}\wedge d\Gamma^{B}_{0}
\end{equation}
( note that $\Omega^{0}(\Gamma_{0})$ in (4.4) and first term in decomposition
(4.3) are the same ) one looks for the solution $\Gamma(\Gamma_{0})$ of the equation
\begin{equation}
\Omega_{AB}(\Gamma(\Gamma_{0}))d\Gamma^{A}(\Gamma_{0})\wedge
d\Gamma^{B}(\Gamma_{0})=\Omega^{0}_{AB}(\Gamma_{0})d\Gamma^{A}_{0}\wedge
d\Gamma^{B}_{0}
\end{equation}
in the form of the power series in the parameter of the charge $e$:
\begin{equation}
\Gamma^{A}=\Gamma^{A}(\Gamma_{0})=\sum_{n=0}^{\infty}e^{n}\Gamma_{n}^{A}(\Gamma_{0}) \end{equation}
Substitution of (4.6) to (4.5) leads to
\begin{equation}
\Omega^{0}_{AC}\partial_{B}\Gamma_{n}^{C}+\Omega^{0}_{CB}\partial_{A}\Gamma_{n}^{C}
+\partial_{C}\Omega^{0}_{AB}\Gamma_{n}^{C}=K_{n}(\Gamma_{1},...,\Gamma_{n-1})
\end{equation}
The last equation may be presented in the form
\begin{equation}
{\cal L}_{\Gamma_{n}}\Omega^{0}=K_{n}(\Gamma_{1},...,\Gamma_{n-1})
\end{equation}
or in a manifest cohomological form
\begin{equation}
d\circ(\Omega^{0}\rule[-0.9mm]{3mm}{.4pt}\!\!\left.\right\rfloor\Gamma_{n})=
K_{n}
\end{equation}
where the operator $d$ stands for an exterior derivation. Then the local
compatibility condition for (4.9) reads as
\begin{equation}
d\circ K_{n}=0
\end{equation}
The condition holds by virtue of Darboux theorem. General solution to the
equation (4.9) has the form
\begin{equation}
\Gamma_{n}=(\Omega_{0}^{-1})^{AB}(\int_{0}^{1}d\tau\tau\Gamma_{0}^{C}K^{n}_{BC}
(\Gamma_{0}\tau)+\partial_{B}\Phi_{n})
\end{equation}
Arbitrary functions $\Phi_{n}$, being involved in the general solution,
does not contribute the respective two-form, hence $\Phi_{n}$  can be
omitted.

Consider smooth structure of the mapping (4.6). On the surface $\Phi =0$,
the 2-form has null-vectors, therefore the Darboux co-ordinates on the map
${\cal M}^{-}$ can not be smoothly continued through $\Phi =0$ to
${\cal M}^{+}$.  Consider the Jacobi matrix
$J^{A}{}_{B}=\partial\Gamma^{A}/\partial\Gamma_{0}^{B}$. From the
equation (4.5)
one may get
\begin{equation}
\det J=\sqrt{\frac{\det\Omega^{0}}{\det\Omega}}, \qquad
\det\Omega^{0}=1, \qquad \det\Omega=\Phi^{2}
\end{equation}
As is seen, the Jacobian $J$ has a discontinuity on the singular surface.
Thus, the perturbative quantisation, being based on the expansion
(4.6) of the Darboux coordinates, is able to reflect the dynamics in single
part (of the phase manifold) alone of the two ones, ${\cal M}^{+}$ or
${\cal M}^{-}$.

%-----------------------------------------------------------------------------
\section{Concluding remarks}

\setcounter{equation}{0}
%-----------------------------------------------------------------------------

In this paper the new understanding of the "consistency" is proposed
for a massive spinning particle's dynamics in the background field.
To have a consistent dynamics, one should satisfy the following
conditions:

1) To provide a consistent deformation of gauge algebra,
   in the conventional sense.

2) To make particle's dynamics smooth upon the singular surface

The last condition results for gyromagnetic ratio to be $g=2$ for the
particle. There is some point of issue in view of the quantization problem.
Existence of the Dirac brackets possessing a {\it local} singularity appears
to be unnoticed before in a realistic constrained system and by virtue of the
fact all the quantization prescriptions demand given Poisson tensor to be
smooth on a whole phase manifold.  It is unclear now how one should proceed
to take the proper account of the singular surface in a quantization scheme.
We drop out the singularity problem in the framework of perturbative approach
to quantization, considering it as a first step towards the globally defined
interacting quantum theory. The classical interaction is seen to be well
defined globally iff $g=2$. Mention that in higher dimensions the
interaction is well defined locally for an arbitrary spin massive particle,
without any restriction to the giromagnetic ratio
\cite{hdim2} although the global analysis, if it was performed for
$d>2+1$ along the lines of this note, would probably result in the same
critical meaning for $g$ in any dimension.

%-----------------------------------------------------------------------------
\section{Appendix}

\setcounter{equation}{0}
%-----------------------------------------------------------------------------

We use realization  of Lobachevsky plane as unit
open disk in a complex plane ${\cal L}\cong \{ z\in C^{1}, z\bar{z}<1\}$.
The proper orthochronous Lorentz group $SO^{\uparrow}(1,2)\cong
SU(1,1)/Z_{2}$ acts on ${\cal L}$ by fractional linear transformations
\begin{equation}
\begin{array}{ccccc} N: &
z\rightarrow z^{\prime}=(az-b)/ (\bar{a}-\bar{b}z)& & & N\in SU(1,1)
\end{array}
\end{equation}
or in infinitesimal form
\begin{equation}
\begin{array}{ccccc}
\delta z = -i\omega_{a}\xi^{a}& & & &\delta\bar{z} =
i\omega_{a}\bar{\xi}^{a}
\end{array}
\end{equation}
where $\omega_{a}$ - small real parameters and $\xi_{a}$ is defined as
follows
\begin{equation}
\xi_{a}=-1/2(2z,1+z^{2},i(z^{2}-1))
\end{equation}
Inner space may be parametrized covariantly by unit timelike
vector representing stereographic map of pseudosphere on Lobachevsky
plane
\begin{equation}
n_{a}=(\frac{1+z\bar{z}}{1-z\bar{z}},\frac{z+\bar{z}}{1-z\bar{z}},i\frac{z-\bar{z}}{1-z\bar{z}})
\end{equation}
The following identity \cite{superanyon} holds for arbitrary 3-vector $s_{a}$
\begin{equation}
s_{a}=2\frac{(s\xi)}{\zeta^{2}}\xi_{a}+2\frac{(s\bar{\xi})}{\zeta^{2}}\bar{\xi}_{a}-(sn)n_{a},
\qquad \zeta=1-z\bar{z}
\end{equation}
This identity allows to expand any 3d vector into fixed values (6.3), (6.4),
that is directly and indirectly used in this note.

%%%%%%%%%%%%%%%%%%%%%%%%%%%%%%%%%%%%%%%%%%%%%%%%%%%%%%%%%%%%%%%%%%%%%%%%%%%%%

\vspace{1.0cm}

{\bf Acknowledgments}

{\small We are thankful to I.A.Bandos, I.A.Batalin, I.V.Gorbunov,
M.Henneaux, A.A.Sharapov, K.M.Shekhter, M.A.Vasiliev for useful discussion
related to various issues concerning the consistency problem for
interactions.}

{\small This work is partially supported by the grant RFBR-98-02-16261,
SLL is partially supported by the grant RFBR-99-01-00980 }

%%%%%%%%%%%%%%%%%%%%            REFERENCES            %%%%%%%%%%%%%%%%%%%%%%

\end{document}